\documentclass[twocolumn]{emulateapj}
\usepackage{graphicx}
\begin{document}

\title{Tidal perturbations to the gravitational inspiral of J0651+2844}

\author{Matthew J Benacquista}

\affil{Center for Gravitational Wave Astronomy}
\affil{University of Texas at Brownsville}
\affil{80 Ft. Brown, Brownsville, TX 78520}
\email{matthew.benacquista@utb.edu}

\begin{abstract}
The recently discovered J0651+2844 is a detached, eclipsing white dwarf binary with an orbital period of 765 s. We investigate the prospects for the detection of gravitational radiation from this system and estimate the effect of the tidal deformation of the low-mass component on the period evolution of the system. Because of the high inclination of the system, the amplitude of the gravitational waves at Earth will be as much as a factor of two lower than that from an optimally oriented system. The dominant contribution of tidal corrections to the period evolution comes from the increase in rotational energy of the components as they spin up to remain tied to the orbital period. This contribution results in an advance of the timing of the eclipses by an additional 0.3 s after one year.
\end{abstract}

\keywords{Binaries: eclipsing---Gravitational waves---White dwarfs}

\section{Introduction}
The Galactic population of white dwarf binaries is expected to produce a foreground of gravitational radiation signals in the millihertz frequency band~\citep{evans87,hils90,hils00,nelemans03,benacquista04,benacquista06,ruiter10}. Most of the known binaries that will be detectable by planned space-based detectors such as the Laser Interferometer Space Antenna (LISA) are interacting systems, and so the period evolution of these systems will be dominated by mass transfer. The recent discovery of SDSS J065133.33+284423.3 (hereafter J0651), a detached eclipsing white dwarf binary with a 12 minute orbital period~\citep{brown11} provides the opportunity for indirect observation of gravitational radiation through the measurement of the period evolution. If detectable by a future detector such as LISA, the combination of direct measurement of gravitational radiation and the observation of the period evolution through timing of the eclipses in the intervening years will allow for detailed study of the internal structure of the component white dwarfs through the dissipation of orbital energy via tidal deformation. In this paper, we evaluate the expected contribution of the tidal and rotational deformation of the components to the period evolution of the binary, and the effect on the direct measurement of gravitational radiation with future detectors.

J0651 is an eclipsing binary with an angle of inclination of $i = 86.9^{+1.6}_{-1.0}$ degrees \citep{brown11}.  They report an orbital period of $P = 765.2062 \pm 0.003$ s for J0651, with component masses of $M_p = 0.25~{\rm M_\odot}$ and $M_s = 0.55~{\rm M_\odot}$. Ellipsoidal variations in the light curve indicate a 5\% difference in the projected area of the primary, showing that it is tidally distorted. The sinusoidal velocity curve indicates that the orbit is circular, while the symmetric appearance of the ellipsoidal variations indicates that the primary is nearly tidally locked. Both the primary and secondary have radii that are slightly larger than those predicted by models of isolated white dwarfs. The primary has an observed radius of $0.0353 \pm 0.0004~{\rm R_\odot}$, which is 5\% larger than models predict~\citep{panei07}. The secondary has an observed radius of $0.0132 \pm 0.0003~{\rm R_\odot}$, which is  4\% greater than models predict. These variations in the sizes and shapes of the components indicate that they are tidally distorted due to their close proximity. This distortion will have an effect on the orbital period evolution. If this effect is not properly taken into account, calculated values of the binary properties from gravitational radiation observations will be in error. Previous studies of the effect of tidal energy dissipation and apsidal precession on detached binaries have focused on  eccentric systems, where the apsidal precession can be measured~\citep{valsecchi11,willems08} or where the varying tidal stresses excite internal modes~\citep{willems10}, or circular systems with non-rotating components~\citep{fuller11a}, or through electromagnetic interactions~\citep{wu08}. Although the components of J0651 appear to be tidally locked, the gradual decay of the orbit due to gravitational radiation requires there to be a slight mismatch between the spin frequency of the components and the orbital frequency of the binary. The slow spin up of the components as a result of the torques applied due to this mismatch will provide the bulk of the deviations from pure gravitational wave emission in the period evolution of the binary.

\section{Orbital Evolution}
The tidal distortions produce non-zero quadrupole moments in the components. If we include the quadrupole moments in the expansion of the potential, we find:
\begin{equation}
V(r) = - \frac{GM\mu}{r}\left[1 + \frac{1}{2}\frac{Q}{Mr^2}\right]
\end{equation}
where $M$ is the total mass of the system, $\mu = M_pM_s/M$ is the reduced mass, and $Q$ is the quadrupole moment. Here, we have assumed that the secondary is not significantly distorted (as will be shown later), and that the quadrupole moment of the primary is symmetric so that it can be described by a single value. This potential results in a variation of Kepler's law, so that the orbital angular frequency ($\omega$) and orbital radius ($a$) are related by:
\begin{equation}
\omega^2 a^3 = GM\left(1+\frac{1}{2}\frac{Q}{Ma^2}\right).
\end{equation}
Since the orbital radius is not directly measurable in gravitational radiation, the orbital angular frequency is the variable of interest. We will show later that the quadrupole correction will be less than 1\%, so we can solve for $a$ as a perturbation about the Keplerian solution, $a_0 = \left(GM/\omega^2\right)^{1/3}$. We set $a = a_0\left(1 + \delta a/a_0\right)$, and find the perturbation to be:
\begin{equation}
\label{deltaa}
\frac{\delta a}{a_0} = \frac{1}{2}\frac{Q\omega^{4/3}}{G^{2/3}M^{5/3}}.
\end{equation}
We are now in a position to calculate the gravitational wave strain and the gravitational wave power for a perturbed system. Assuming point masses for the orbital configuration in order to calculate the total quadrupole moment of the binary system following the method of~\citet{peters63}, we find that the power radiated through gravitational radiation is:
\begin{eqnarray}
P_{\rm GW} & = &  \frac{32}{5}\frac{G}{c^5}\mu^2a^4\omega^6 \nonumber \\
& = & \frac{32}{5}\frac{G^{7/3}}{c^5}\mu^2M^{4/3}\omega^{10/3}\left(1 + 2\frac{Q\omega^{4/3}}{G^{2/3}M^{5/3}}\right).
\end{eqnarray}

The energy radiated by gravitational waves comes from the orbital energy of the system, and this manifests itself in a gradual shrinking of the orbit with an increase in the orbital frequency, known as the ``chirp.'' In the case of J0651, the tidal distortion of the components provides a torque that spins up the components so that their rotations remain locked to the orbital period. The increase in the rotational kinetic energy of the components also comes from the orbital energy of the system, and so the increase in orbital frequency will be larger than that solely due to the emission of gravitational radiation. 

If the torques on the tidal deformations couple to any internal modes for each component, additional energy can be lost to heating of the component stars. However, the resonance condition for coupling the tidal deformations to these internal modes requires that the driving frequency of the distortion be comparable to the natural frequency of one of the modes. The driving frequency is given by:
\begin{equation}
\omega_{t} = 2\left(\omega-\Omega\right)
\end{equation}
where $\omega$ is the orbital frequency and $\Omega$ is the spin frequency of the white dwarf~\citep{fuller11b}. Since the lowest normal mode frequency in a white dwarf is on the order of a mHz~\citep{winget08}, and the orbital frequency is 1.3 mHz, then $\Omega\le\omega/2$ in order for $\omega_{t}$ to be comparable to any normal mode frequency. This would be clearly visible in the light curve, and so we can neglect coupling to the internal modes. An additional channel for energy loss in the system may be through heating without any direct coupling to normal modes. \citet{fuller11b} have shown that the tidal heating rate can be compared to the tidal spin-up energy rate by:
\begin{equation}
\dot{E}_{\rm heat} = \dot{E}_{\rm tide}\left(1-\frac{\Omega}{\omega}\right),
\end{equation}
and so if $\Omega\sim\omega$, then the energy loss due to heating can also be neglected. A more detailed treatment of the effects of tidal heating on the spin evolution of J0651 and the implications of a non-negligible contribution to heating can be found in~\citet{piro11}.

Here, we obtain a lower bound on the increase in the chirp by only considering the rotational kinetic energy and ignoring any internal heating due to coupling with internal modes, thus the change in total energy of the system is equal to the power radiated through gravitational radiation, so:
\begin{equation}
-P_{\rm GW} = \frac{d}{dt}\left[\frac{1}{2}\left(\mu a^2 + I_p + I_s\right)\omega^2 + V(a)\right]
\end{equation}
where $I_p$ and $I_s$ are the moments of inertia of the primary and the secondary, respectively. Introducing the dimensionless quantities due to the quadrupole correction to the potential ($\Delta_{\rm Q}$) and the spin-up of the components ($\Delta_{\rm I}$), where:
\begin{eqnarray}
\Delta_{\rm Q} & = & \frac{Q\omega^{4/3}}{G^{2/3}M^{5/3}}\label{dq} \\
\Delta_{\rm I} & = & \frac{\left(I_p + I_s\right)\omega^{4/3}}{\mu G^{2/3}M^{2/3}}\label{dI},
\end{eqnarray}
we find
\begin{equation}
\dot{\omega} =\dot{ \omega}_0\left(1 + 5\Delta_{\rm Q} + 3\Delta_{\rm I}\right)
\end{equation}
where $\dot{\omega}_0$ is the value of the chirp in the absence of any tidal deformations:
\begin{equation}
\dot{\omega}_0 = \frac{96}{5}\frac{G^{5/3}}{c^5}\mu M^{2/3}\omega^{11/3}.
\end{equation}
The chirp can be used in combination with the gravitational wave frequency ($\omega_{\rm GW} = 2\omega$) to determine the ``chirp mass'', a combination of the reduced mass and the total mass given by: ${\mathcal M}_c = \mu^{3/5}M^{2/5}$. If the corrections due to tidal deformation are not properly accounted for, the calculated chirp mass ($\tilde{\mathcal M}_c$) will be greater than the true chirp mass by:
\begin{equation}
\label{chirpmass}
\tilde{\mathcal M}_c = \left(\frac{5c^5\dot{\omega}}{96G^{5/3}\omega^{11/3}}\right)^{3/5} = {\mathcal M}_c\left(1 + 3\Delta_{\rm Q} +\frac{9}{5}\Delta_{\rm I}\right).
\end{equation}

The amplitude of the two polarizations are given by~\citep[e.g.:][]{cutler98} as:
\begin{eqnarray}
A_+ & = & \frac{2GM_pM_s}{c^4ad}\left(1+\cos^2{i}\right)\\
A_\times & = & -\frac{4GM_pM_s}{c^4ad}\cos{i},
\end{eqnarray}
where $d$ is the distance to the binary and $i$ is the inclination. We note that the fact that J0651 is eclipsing implies $\cos{i} \simeq 0.05$ and so the $\times$-polarization amplitude is near zero and the $+$-polarization amplitude is near its minimum. Thus, the strain amplitude may be as much as a factor of two lower than the value estimated in~\citet{brown11}.  Taking into account the perturbation in $a$ due to the deformation of the components (see Equation~\ref{deltaa}), we can express the basic amplitude as:
\begin{equation}
\label{strainamp}
h = \frac{G^2M_pM_s}{c^4ad} = \frac{G^{5/3}\mu M^{2/3}\omega^{2/3}}{c^4 d}\left(1 + \Delta_{\rm Q}\right).
\end{equation}

Using the chirp mass calculated from $\omega$ and $\dot{\omega}$, to eliminate $\mu M^{2/3}$ from $h$,~\citet{schutz97} showed that the distance to the binary can be calculated using:
\begin{equation}
\label{schutzdistance}
d = \frac{5c\dot{\omega}}{96h\omega^3}.
\end{equation}
However, this expression is only valid for point masses and assumes that the chirp is entirely due to energy loss through gravitational radiation. Including effects due to the tidal and rotational distortions of the component masses results in the following equation for the distance:
\begin{equation}
d = \frac{5c\dot{\omega}}{96h\omega^3}\left(1 + 6\Delta_{\rm Q} + 3\Delta_{\rm I}\right).
\end{equation}
Thus, we see that including effects due to tidal deformation yields corrections to Kepler's law, and calculated values of the chirp, chirp mass, and distance to the binary. All of these corrections can be expressed in terms of $\Delta_{\rm Q}$ and $\Delta_{\rm I}$. In the next section, we estimate the size of these corrections in J0651.

\section{Perturbations in J0651}
Both components of J0651 are relatively cool white dwarfs with masses substantially below the Chandrasekhar mass. Consequently, we can consider them to be $n = 1.5$ polytropes. \citet{sirotkin09} looked at perturbations in the internal structure for polytropes in close binaries using a self-consistent field method to calculate the effects of rotation and tidal deformation in synchronized, circular-orbit binaries. Their calculations assume that one component of the binary is a point mass and they then determine the perturbations of the other component. The volume radius $R$ is defined as the radius of a sphere with the equivalent volume of the distorted polytrope, and the dimensionless volume radius ($\xi_1$) is determined by:
\begin{equation}
R = \alpha\xi_1 = \left(\frac{3V}{4\pi}\right)^{1/3}
\end{equation}
where
\begin{equation}
\alpha = \left[\frac{5K}{8\pi G}\rho_c^{1/3}\right]^{1/2}
\end{equation}
and $K$ and $\rho_c$ are the pressure constant and the central density, respectively. The mean density is determined by the mass ($M_*$) of the star and the volume radius, so that $\bar{\rho} = 3M_*/\left(4 \pi R^3\right)$. A dimensionless description of the mass concentration is given by $\rho_c/\bar{\rho}$. The moment of inertia ($I$), is characterized by the dimensionless quantity:
\begin{equation}
\tilde{I} = \frac{1}{\alpha^5\rho_c}\int{\rho\left(x^2 + y^2\right)d^3x},
\end{equation}
where the orbital plane is assumed to lie in the $xy$-plane. \citet{sirotkin09} produce numerical fitting formulae for $\xi_1$, $\tilde{I}$, and $\rho_c/\bar{\rho}$ in terms of the relative radius $R_A = R/a$. We can recover the moment of inertia from the dimensionless quantities, using $R_A$,  $a$, and $M_*$:
\begin{equation}
I = \frac{R_A^2}{\xi_1^5}\left(\frac{\rho_c}{\bar{\rho}}\right)\tilde{I}\left(\frac{3}{4\pi}\right)M_*a^2.
\end{equation}
Since any corrections to the orbital separations will result in second order corrections to the orbital evolution, we assume that the orbital separation is the unperturbed separation for these calculations, and so $a = 0.1676~{\rm R_\odot}$. 

For the primary, we have $M_p = 0.25~{\rm M_\odot}$, $R_A = 0.21$, and the mass fraction $M_p/M = 0.3$. Using Tables~4, 5, and 7 from~\citet{sirotkin09}, we find $\rho_c/\bar{\rho} = 6.0486$, $\xi_1 = 3.6783$, and $\tilde{I} = 92.0131$. This corresponds to a moment of inertia for the primary of $I_p = 6.111\times10^{-5}~{\rm M_\odot R_\odot^2} = 5.912\times10^{43}~{\rm kg\cdot m^2}$. For the secondary, we have $M_s = 0.55~{\rm M_\odot}$, $R_A = 0.08$, and the mass fraction $M_s/M = 0.7$. This gives $\rho_c/\bar{\rho} = 5.9901$, $\xi_1 = 3.6538$, and $\tilde{I} = 90.9334$, corresponding to a moment of inertia for the secondary of $I_s = 1.914\times10^{-5}~{\rm M_\odot R_\odot^2} = 1.851\times10^{43}~{\rm kg\cdot m^2}$. These values, along with those for an unperturbed, isolated $n=1.5$ polytrope are given in Table~\ref{wdstructures}. From here we can see that the secondary is nearly identical to an unperturbed white dwarf with the exception of a lower moment of inertia. We can interpret this as implying that the secondary is relatively undistorted. Thus, we need only consider the quadrupole moment of the primary.

\begin{deluxetable}{lrrr}
\tablecaption{Structure parameters of J0651 components and unperturbed $n=1.5$ polytropes.\label{wdstructures}}
\tablehead{ & \colhead{$\rho_c/\bar{\rho}$} & \colhead{$\xi_1$} & \colhead{$\tilde{I}$}}
\startdata
Primary & 6.0486 & 3.6783 & 92.0131 \\
Secondary & 5.9901 & 3.6538 & 90.9334 \\
Unperturbed & 5.9907 & 3.6538 & 93.156
\enddata
\end{deluxetable}

In order to estimate the quadrupole moment of the primary, we note that the quadrupole moment is proportional to the moment of inertia with a proportionality constant related to the structure of the star. In this case, we do not have a detailed description of the structure of the primary, but we can estimate an upper bound on the quadrupole moment by assuming constant density and a shape that deviates from spherical by a small amount. Since the light curve of J0651 indicates a 5\% ellipsoidal variation of the primary. Thus, we assume a 5\% deviation from spherical and compute the quadrupole moment. It can be shown that to first order in the deviation, the quadrupole moment is 5\% of the moment of inertia. Therefore, we take as an upper bound, $Q = 0.05I_p = 2.956\times10^{42}~{\rm kg\cdot m^2}$.

Taking the above values for $I_p$, $I_s$, and $Q$ and using the observed values of the masses and frequency, we find that the dimensionless perturbations of Equations~\ref{dq} and~\ref{dI} are:
\begin{eqnarray}
\Delta_{\rm Q} & = & 1.36\times10^{-4} \\
\Delta_{\rm I} & = & 0.0166.
\end{eqnarray}

\section{Discussion}
The perturbations introduced by the distortion and tidal locking of the components of J0651 will be on the order of a few percent. Nearly all of this will come from the spin-up of the components due to the torques induced through the inspiral in order to maintain tidal locking. The effects of a non-zero quadrupole moment will contribute less than 0.1\% to any perturbation, and can safely be ignored at the separations found in J0651. If the effect of spin-up is ignored in future gravitational observations, then the chirp mass will be over-estimated by 3\%, which is well within the errors expected from spectrum modeling. The distances will then also be over-estimated if the standard formula of~\citet{schutz97} is used. The calculated distance will be nearly 5\% greater than the true distance. At a current estimated distance of 1~kpc, and an apparent magnitude of $g = 19.1$~\citep{brown11}, GAIA can be expected to provide a distance accuracy of around 8-20\%~\citep{bailerjones09}. Thus, ignoring the perturbations due to tidal distortions and tidal locking when using gravitational wave observations will still provide better estimates of the distance and masses of white dwarf binaries such as J0651 than can be obtained through electromagnetic observations.

One area in which electromagnetic observations will be superior to gravitational wave observations will be in the measurement of the chirp. An estimate of the minimum measurable chirp through gravitational wave observations can be found by requiring the gravitational wave frequency of the binary to shift by one resolvable frequency bin during the observation time $T_{\rm obs}$. Since the size of the resolvable frequency bins scales as the inverse of the observation time, the minimum measurable chirp is estimated by $\dot{f_{\rm GW}} \ge T_{\rm obs}^{-2}$, where $f_{\rm GW} = \omega/\pi$. Ignoring the perturbations, $\dot{\omega} = 9.184\times10^{-17}~{\rm Hz}^2$, which implies a 6 year observation time. The large signal-to-noise ratio expected from this source, combined with improved data analysis may shorten this requirement to two or three years. On the other hand, timing of the eclipses allows for the chirp to measured within one year. After a time interval $T$, the eclipses will occur earlier by an amount:
\begin{equation}
\delta t = \frac{1}{2} \frac{\dot{\omega} T^2}{\omega + \dot{\omega}T}.
\end{equation}
If the perturbations are ignored, $\delta t = 5.57~{\rm s}$ for $T = 1~{\rm yr}$. Including the perturbations  shifts this to $\delta t = 5.85$ s --- a difference of less than 0.3 s. However, after two years, this difference grows to more than 1 s.

Combining both electromagnetic and gravitational wave observations will allow for the properties of J0651 to be measured to better precision than either observation alone can achieve. Assuming that a space-based observatory such as LISA is launched in the early 2020's, the timing of the eclipses of J0651 will have measured an accumulated advance of the eclipse of nearly 10 minutes, 30 seconds of which can be attributed to tidal effects. Thus, estimates of the distance to J0651 can be made using the gravitational wave chirp as measured by eclipses and the strain amplitude as measured by LISA without having to wait for the chirp to be measured in the gravitational wave signal. Furthermore, astrometric measurements of J0651s distance will have been made by GAIA, and this can be compared with the gravitational wave distance. Keeping in mind that the perturbations due to the tidal interactions that we have calculated here ignore any thermal dissipation of energy into heating of the white dwarfs, the true perturbation may be large enough to cause a discrepancy between the gravitational wave distance and the astrometric distance.

\section{Conclusion}
We have shown that the tidal deformation of the components of J0651 will cause the orbital evolution to deviate from that predicted by the simple loss of energy due to the emission of gravitational radiation as some orbital energy will go into spinning up the components in order to maintain tidal locking. Ignoring any additional dissipation of energy through heating of the component white dwarfs, this extra energy sink will cause the chirp to be about 5\% greater than due to gravitational radiation alone. This is a lower bound on the perturbation, as thermal dissipation will only extract more energy from the orbit. Although the deformation will also cause the orbit to deviate from pure Keplerian, we have shown that this effect will contribute less than 0.1\% to the chirp, and can probably be safely ignored.

As J0651 is an eclipsing system, its inclination is quite high. This has a disadvantage for gravitational wave observations, since the gravitational wave strain is more than a factor of two weaker for this orientation compared with the optimal orientation of face-on. On the other hand, timing of the eclipses allows for the chirp to measured to better accuracy over a shorter observing time than can be done with gravitational wave observations. Since the precision of this measurement increases with observing time and no gravitational wave observatory in this frequency band is expected for at least a decade, eclipse timing will provide an exquisitely measured chirp for use with gravitational wave observations as soon as the detectors come online. In the intervening years between now and the launch of millihertz gravitational wave observatories, GAIA will have completed its mission and analyzed parallax measurements for J0651 (if it is targeted). The distance obtained through parallax is expected to have errors of 8-20\%, while the greatest errors in the distance obtained through gravitational waves will be due to the uncertainties in the amount of energy dissipated through tidal interactions. We have estimated this to be smaller than the errors in parallax measurements. Thus, millihertz gravitational wave observations will provide better measurements of the distances to detached double white dwarfs such as J0651 and other faint objects than can be obtained through parallax measurements, with the dominant source of error being the tidal interactions.

\acknowledgements
This work was supported by the Center for Gravitational Wave Astronomy, partially funded through NASA grant NNX09AV06A. I would like to thank the anonymous referee whose suggestions have helped clarify the assumptions in this work, and Francesca Valsecchi and Mukremin Kilic for helpful discussions.

\end{document}